# TOWARDS A GUIDELINE FOR EVALUATION METRICS IN MEDICAL IMAGE SEGMENTATION

*Dominik Müller[1,2], Iñaki Soto-Rey[2] and Frank Kramer[1]*

[1]IT-Infrastructure for Translational Medical Research, University of Augsburg, Germany
[2]Medical Data Integration Center, Institute for Digital Medicine, University Hospital Augsburg, Germany

## ABSTRACT

**In the last decade, research on artificial intelligence has seen rapid growth with deep learning models, especially in the field of medical image segmentation. Various studies demonstrated that these models have powerful prediction capabilities and achieved similar results as clinicians. However, recent studies revealed that the evaluation in image segmentation studies lacks reliable model performance assessment and showed statistical bias by incorrect metric implementation or usage. Thus, this work provides an overview and interpretation guide on the following metrics for medical image segmentation evaluation in binary as well as multi-class problems: Dice similarity coefficient, Jaccard, Sensitivity, Specificity, Rand index, ROC curves, Cohen's Kappa, and Hausdorff distance. As a summary, we propose a guideline for standardized medical image segmentation evaluation to improve evaluation quality, reproducibility, and comparability in the research field.**

## INTRODUCTION

In the last decade, research on artificial intelligence has seen rapid growth with deep learning models, by which various computer vision tasks got successfully automated through accurate neural network classifiers [1]. Evaluation procedures or quality of model performance are highly distinctive in computer vision between different research fields and applications.

The subfield medical image segmentation (MIS) covers the automated identification and annotation of medical regions of interest (ROI) like organs or medical abnormalities (e.g. cancer or lesions) [2]. Various novel studies demonstrated that MIS models based on deep learning revealed powerful prediction capabilities and achieved similar results as radiologists regarding performance [1,2]. Clinicians, especially from radiology and pathology, strive to integrate deep learning based MIS methods as clinical decision support (CDS) systems in their clinical routine to aid in diagnosis, treatment, risk assessment, and reduction of time-consuming inspection processes [1,2]. Throughout their direct impact on diagnosis and treatment decisions, correct and robust evaluation of MIS algorithms is crucial.

However, in the past years a strong trend of highlighting or cherry-picking improper metrics to show particularly high scores close to 100% was revealed in scientific publishing of MIS studies [3–7]. Studies showed that statistical bias in evaluation is caused by issues reaching from incorrect metric implementation or usage to missing hold-out set sampling for reliable validation [3–11]. This led to the current situation that various clinical research teams are reporting issues on model usability outside of research environments [4,7,12–16]. The use of faulty metrics and missing evaluation standards in the scientific community for the assessment of model performance on health-sensitive procedures is a large threat to the quality and reliability of CDS systems.

In this work, we want to provide an overview on appropriate metrics, demonstrate their interpretation, and propose a guideline for properly evaluating medical image segmentation performance in order to increase research reliability and reproducibility in the field of medical image segmentation.

## MAIN TEXT

### Evaluation Metrics

Evaluation of semantic segmentation can be quite complex because it is required to measure classification accuracy as well as localization correctness. The aim is to score the similarity between the predicted (prediction) and annotated segmentation (ground truth). Over the last 30 years, a large variety of evaluation metrics can be found in the MIS literature [10]. However, only a handful of scores have proven to be appropriate and are used in a standardized way [10]. The behavior of all presented metrics in this work is illustrated in Figure 1 which summarizes the metric application on multiple use cases. In the following chapters, each metric will be defined and discussed in terms of possible issues. Nearly all presented metrics, except Hausdorff distance, are based on the computation of a confusion matrix for a binary segmentation task, which contains the number of true positive (TP), false positive (FP), true negative (TN), and false negative (FN) predictions. Except for Cohen's Kappa and Hausdorff distance, the value ranges of all presented metrics span from zero (worst) to one (best).

## F-measure based Metrics

F-measure, also called F-score, based metrics are one of the most widespread scores for performance measuring in computer vision as well as in the MIS scientific field [10,11,17,18]. It is calculated from the sensitivity and precision of a prediction, by which it scores the overlap between predicted segmentation and ground truth. Still, by including the precision, it also penalizes false positives, which is a common factor in highly class imbalanced datasets like MIS [10,11]. Based on the F-measure, there are two popular utilized metrics in MIS: The Intersection-over-Union (IoU), also known as Jaccard index or Jaccard similarity coefficient, and the Dice similarity coefficient (DSC), also known as F1 score or Sørensen-Dice index. Besides that the DSC is defined as the harmonic mean between sensitivity and precision, the difference between the two metrics is that the IoU penalizes under- and over-segmentation more than the DSC. Even so, both scores are appropriate metrics, the DSC is the most used metric in the large majority of scientific publications for MIS evaluation [10,11,18].

$$IoU = \frac{TP}{TP + FP + FN} \qquad (1)$$

$$DSC = \frac{2TP}{2TP + FP + FN} \qquad (2)$$

## Sensitivity and Specificity

Especially in medicine, specificity and sensitivity are established standard metrics for performance evaluation [10,11]. For pixel classification, the sensitivity (Sens), also known as recall or true positive rate, focuses on the true positive detection capabilities, whereas the specificity (Spec), also known as true negative rate, evaluates the capabilities for correctly identifying true negative classes (like the background class). In MIS evaluation, the sensitivity is a valid and popular metric, but still less sensitive to F-score based metrics for exact evaluation and comparison of methods [10,11]. However, the specificity can result in an improper segmentation metric if not correctly understood. In MIS tasks, the specificity indicates the model's capability to detect the background class in an image. Due to the large fraction of pixels annotated as background compared to the ROI, specificity ranges close to 1 are standard and expected. Thus, specificity is a suited metric for ensuring model functionality, but less for model performance.

$$Sensitivity = \frac{TP}{TP + FN} \qquad (3)$$

$$Specificity = \frac{TN}{TN + FP} \qquad (4)$$

## Accuracy / Rand Index

Accuracy (Acc), also known as Rand index or pixel accuracy, is one or even the most known evaluation metric in statistics [10]. It is defined as the number of correct predictions, consisting of correct positive and negative predictions, compared to the total number of predictions. However, it is strongly discouraged to use accuracy in MIS. The reason for this is that medical segmentation images are commonly highly class imbalanced [10,19]. A medical image usually contains a single ROI taking only a small percentage of pixels in the image, whereas the remaining image is all annotated as background. Because of the true negative inclusion, the accuracy metric will always result in an illegitimate high scoring. Even predicting the segmentation of an entire image as background class, accuracy scores are often higher than 90% or even close to 100%. Therefore, the misleading accuracy metric is not suited for MIS evaluation and using it is highly discouraged in scientific evaluations.

$$Accuracy = \frac{TP + TN}{TP + TN + FN + FP} \qquad (5)$$

## Receiver Operating Characteristic

The ROC curve, short for Receiver Operating Characteristic, is a line plot of the diagnostic ability of a classifier by visualizing its performance with different discrimination thresholds [10]. The performance is shown through the true positive rate (TPR) against the false positive rate (FPR). In particular, ROC curves are widely established as a standard metric for comparing multiple classifiers and in the medical field for evaluating diagnostic tests as well as clinical trials [20]. As a single-value performance metric, the area under the ROC curve (AUC) was first introduced by Hanley and McNeil 1982 for diagnostic radiology [21]. Nowadays, the AUC metric is also a common method for the validation of machine learning classifiers. It has to be noted that an AUC value of 0.5 can be interpreted as a random classifier. The following AUC formula is defined as the area of the trapezoid according to David Powers [6]:

$$AUC = 1 - \frac{1}{2}\left(\frac{FP}{FP + TN} + \frac{FN}{FN + TP}\right) \qquad (6)$$

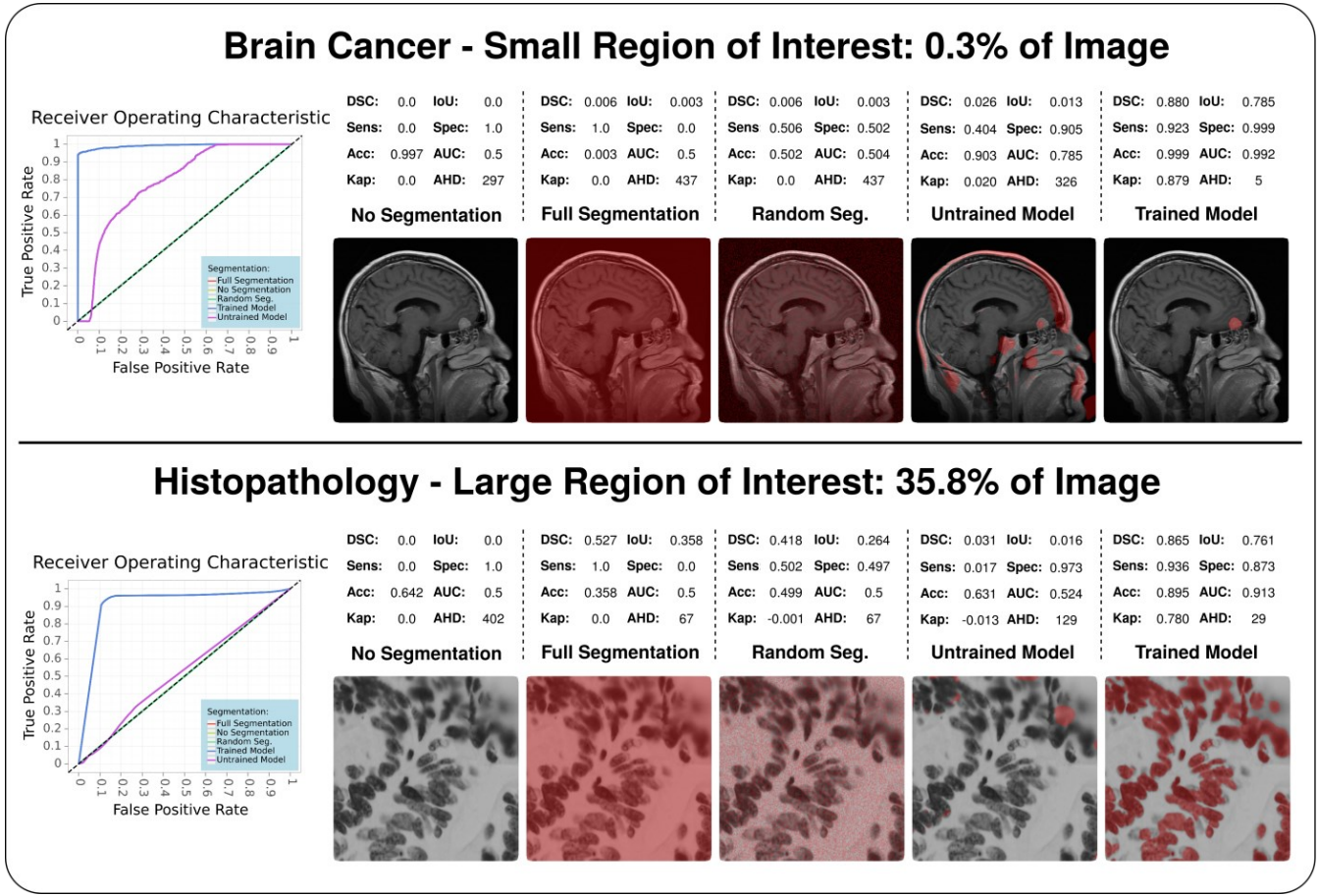

**Figure 1:** Demonstration of metric behavior in the context of different-sized ROIs compared to the total image. The figure is showing the perks of F-measure based metrics like DSC as well as IoU and the inferiority of Rand index usage. Furthermore, the small ROI segmentation points out that metrics like accuracy have no value for interpretation in these scenarios, whereas the large ROI segmentation indicates that small percentage variance can lead to a risk of missing whole instances of ROIs. The analysis was performed in the following scenarios and common MIS use cases. Scenarios: No segmentation (no pixel is annotated as ROI), full segmentation (all pixels are annotated as ROI), random segmentation (full random-based annotation), untrained (after 1 epoch during training) and trained model (fully fitted model). Use cases: Small ROIs via brain tumor detection in magnetic resonance imaging and large ROIs via cell nuclei detection in pathology microscopy. Data and applied method details are described in the "Availability of data and materials" section.

### Cohen's Kappa

The metric Cohen's Kappa (Kap), introduced by Cohen in 1960 in the field of psychology, is a change-corrected measure of agreement between annotated and predicted classifications [10,22,23]. For interpretation, Kap measures the agreement caused by chance like the AUC score and ranges from -1 (worst) to +1 (best), whereas a Kap of 0 indicates a random classifier. Through its capability of application on imbalanced datasets, it has gained popularity in the field of machine learning [23]. However, a recent study demonstrated that it still correlates strongly to higher values on balanced datasets [23,24]. Additionally, it does not allow comparability on different sampled datasets or interpretation on prediction accuracy.

$$f_c = \frac{(TN+FN)(TN+FP)+(FP+TP)(FN+TP)}{TP+TN+FN+FP} \tag{7}$$

$$Kap = \frac{(TP+TN)-f_c}{(TP+TN+FN+FP)-f_c} \tag{8}$$

*Average Hausdorff Distance*

In contrast to other confusion matrix based metrics, the Hausdorff distance (HD) is a spatial distance based metric [10]. The HD measures the distance between two sets of points, like ground truth and predicted segmentation, and allows scoring localization similarity by focusing on boundary delineation (contour) [10,25,26]. Especially in more complex and granular segmentation tasks, exact contour prediction is highly important which is why HD based evaluations have become popular in the field of MIS [10]. However, because the HD is sensitive to outliers, the symmetric Average Hausdorff Distance (AHD) is utilized in the majority of applications instead [10,19,25]. The symmetric AHD is defined by the maximum between the directed average Hausdorff distance d(A,B) and its reverse direction d(B,A) in which A and B represent the ground truth and predicted segmentation, respectively, and ||a-b|| represents a distance function like Euclidean distance [10]:

$$d(A,B) = \frac{1}{N}\sum_{a \in A}\min_{b \in B}||a-b|| \tag{9}$$

$$AHD(A,B) = \max(d(A,B),d(B,A)) \tag{10}$$

*Sample Visualization*

Besides the exact performance evaluation via metrics, it is strongly recommended to additionally visualize segmentation results. Comparing annotated and predicted segmentation allows robust performance estimation by eye. Sample visualization can be achieved via binary visualization of each class (black and white) or via utilizing transparent color application based on pixel classes on the original image. The strongest advantage of sample visualization is that statistical bias, overestimation of predictive power through unsuited or incorrect computed metrics, is avoided.

*Other Metrics*

In the field of MIS, various other metrics exist and can be applied depending on the research question and interpretation focus of the study. This work focused on the most suited metrics to establish a standardized MIS evaluation procedure and to increase reproducibility. For further insights on the theory of previously presented metrics or a large overview of all metrics for MIS, we refer to the excellent studies of Taha et al. [10]. Additionally, Nai et al. provided a high-quality demonstration of various metrics on a prostate MRI dataset [19].

## Multi-Class Evaluation

The previous evaluation metrics discussed are all defined for binary segmentation problems. It is needed to be aware that applying binary metrics on multi-class problems can result in highly biased results, especially in the presence of class imbalance [6]. This can often lead to a confirmation bias and promising-looking evaluation results in scientific publications which, however, are actually quite weak [6]. In order to evaluate multi-class tasks, it is required to compute and analyze the metrics individually for each class. Distinct evaluation for each class is in the majority of cases the most informative and comparable method. Nevertheless, it is often necessary to combine the individual class scores to a single value for improving clarity or for further utilization, for example as loss function. This can be achieved by micro and macro averaging the individual class scores. Whereas macro-averaging computes the individual class metrics independently and just averages the results, micro-averaging aggregates the contributions of each class for computing the average score.

## Evaluation Guideline

➢ Use DSC as main metric for validation and performance interpretation.
➢ Use AHD for interpretation on point position sensitivity (contour) if needed.
➢ Avoid any interpretations based on high pixel accuracy scores.
➢ Provide next to DSC also IoU, Sensitivity, and Specificity for method comparability.
➢ Provide sample visualizations, comparing the annotated and predicted segmentation, for visual evaluation as well as to avoid statistical bias.
➢ Avoid cherry-picking high-scoring samples.
➢ Provide histograms or box plots showing the scoring distribution across the dataset.

➢ For multi-class problems, provide metric computations for each class individually.
➢ Avoid confirmation bias through macro-averaging classes which is pushing scores via background class inclusion.
➢ Provide access to evaluation scripts and results with journal data services or third-party services like GitHub [27] and Zenodo [28] for easier reproducibility.

## OUTLOOK

This work focused on the most suited metrics to establish a standardized medical image segmentation evaluation procedure. We hope that our guidelines will help improve evaluation quality, reproducibility, and comparability in future studies in the field of medical image segmentation. Furthermore, we noticed that there is no universal Python package for metric computations, which is why we are currently working on a package to compute metrics scores in a standardized way. In the future, we want further contribute and expand our guidelines for reliable medical image segmentation evaluation.

## LIST OF ABBREVIATIONS

MIS: Medical Image Segmentation; ROI: Region of Interest; CDS: Clinical Decision Support; TP: True Positive; FP: False Positive; TN: True Negative; FN: False Negative; DSC: Dice Similarity Coefficient; IoU: Intersection-over-Union; Sens: Sensitivity; Spec: Specificity; Acc: Accuracy; ROC: Receiver Operating Characteristic; TPR: True Positive Rate; FPR: False Positive Rate; Kap: Cohen's Kappa; HD: Hausdorff distance; AHD: Average Hausdorff Distance.

## DECLARATIONS

### Ethics approval and consent to participate
Not applicable.

### Consent for publication
Not applicable.

### Availability of data and materials
The analysis utilized the medical image segmentation framework MIScnn [29] and was performed with the following parameters: Sampling in 80% training, 20% testing sets; resizing into 512x512 images; value intensity normalization via Z-score; extensive online image augmentation during training, common U-Net architecture as neural network with focal Tversky loss function and a batch of 24; advanced training features like dynamic learning rate, early stopping and model checkpoints. The training was performed for a maximum of 1000 epochs (77 up to 173 epochs after early stopping) and on 75 randomly selected images per epoch.

The following datasets were used: Brain tumor detection in magnetic resonance imaging from Cheng et al. [30,31] and cell nuclei detection in pathology microscopy from Caicedo et al. [32].

In order to ensure full reproducibility, the complete code of the analysis is available in the following public Git repository:
https://github.com/frankkramer-lab/miseval.analysis
Furthermore, the trained models, evaluation results, and metadata are available in the following public Zenodo repository:
https://doi.org/10.5281/zenodo.5877797

We are currently working on a universal Python package for metric computation: "MISeval: a metric library for Medical Image Segmentation Evaluation". The current work-in-progress version of the package is available in the following public Git repository:
https://github.com/frankkramer-lab/miseval

### Competing interests
The authors declare no conflicts of interest.

### Funding
This work is a part of the DIFUTURE project funded by the German Ministry of Education and Research (Bundesministerium für Bildung und Forschung, BMBF) grant FKZ01ZZ1804E.

### Authors' contributions
FK was in charge of as well as ISR contributed to reviewing, and correcting the manuscript. DM performed the data analysis, and was in charge of manuscript drafting and revision. All the authors are accountable for the integrity of this work. All authors read and approved the final manuscript.

### Acknowledgments
We want to thank Dennis Hartmann, Philip Meyer, Natalia Ortmann and Peter Parys for their useful comments and support.

## REFERENCES

1. Litjens G, Kooi T, Bejnordi BE, Setio AAA, Ciompi F, Ghafoorian M, et al. A survey on deep learning in medical image analysis. Med Image Anal. 2017;42(December 2012):60–88.
2. Müller D, Soto-Rey I, Kramer F. Robust chest CT image segmentation of COVID-19 lung infection based on limited

data. Informatics Med Unlocked [Internet]. 2021 Jan 1 [cited 2021 Jul 29];25:100681. Available from: https://linkinghub.elsevier.com/retrieve/pii/S23529148210 01660

3. Renard F, Guedria S, Palma N De, Vuillerme N. Variability and reproducibility in deep learning for medical image segmentation. Sci Rep [Internet]. 2020 Dec 1 [cited 2022 Jan 8];10(1):1–16. Available from: www.nature.com/scientificreports

4. Parikh RB, Teeple S, Navathe AS. Addressing Bias in Artificial Intelligence in Health Care [Internet]. Vol. 322, JAMA - Journal of the American Medical Association. American Medical Association; 2019 [cited 2022 Jan 8]. p. 2377–8. Available from: https://jamanetwork.com/

5. Zhang Y, Mehta S, Caspi A. Rethinking Semantic Segmentation Evaluation for Explainability and Model Selection.

6. Powers DMW. Evaluation: from precision, recall and F-measure to ROC, informedness, markedness and correlation. 2020 Oct 10 [cited 2022 Jan 8]; Available from: http://arxiv.org/abs/2010.16061

7. El Naqa IM, Hu Q, Chen W, Li H, Fuhrman JD, Gorre N, et al. Lessons learned in transitioning to AI in the medical imaging of COVID-19. J Med Imaging [Internet]. 2021 Oct 1 [cited 2022 Jan 8];8(S1):010902. Available from: https://www.spiedigitallibrary.org/terms-of-use

8. Gibson E, Hu Y, Huisman HJ, Barratt DC. Designing image segmentation studies: Statistical power, sample size and reference standard quality. Med Image Anal. 2017 Dec 1;42:44–59.

9. Niessen WJ, Bouma CJ, Vincken KL, Viergever MA. Error Metrics for Quantitative Evaluation of Medical Image Segmentation. In Springer, Dordrecht; 2000 [cited 2022 Jan 8]. p. 275–84. Available from: https://link.springer.com/chapter/10.1007/978-94-015-9538-4_22

10. Taha AA, Hanbury A. Metrics for evaluating 3D medical image segmentation: Analysis, selection, and tool. BMC Med Imaging [Internet]. 2015 Aug 12 [cited 2021 May 14];15(1):29. Available from: http://bmcmedimaging.biomedcentral.com/articles/10.118 6/s12880-015-0068-x

11. Popovic A, de la Fuente M, Engelhardt M, Radermacher K. Statistical validation metric for accuracy assessment in medical image segmentation. Int J Comput Assist Radiol Surg [Internet]. 2007 Jul 28 [cited 2021 Jan 13];2(3–4):169–81. Available from: https://link.springer.com/article/10.1007/s11548-007-0125-1

12. Sandeep Kumar E, Satya Jayadev P. Deep Learning for Clinical Decision Support Systems: A Review from the Panorama of Smart Healthcare. In Springer, Cham; 2020 [cited 2022 Jan 7]. p. 79–99. Available from: https://link.springer.com/chapter/10.1007/978-3-030-33966-1_5

13. Altaf F, Islam SMS, Akhtar N, Janjua NK. Going deep in medical image analysis: Concepts, methods, challenges, and future directions. Vol. 7, IEEE Access. Institute of Electrical and Electronics Engineers Inc.; 2019. p. 99540–72.

14. Shaikh F, Dehmeshki J, Bisdas S, Roettger-Dupont D, Kubassova O, Aziz M, et al. Artificial Intelligence-Based Clinical Decision Support Systems Using Advanced Medical Imaging and Radiomics. Curr Probl Diagn Radiol. 2021 Mar 1;50(2):262–7.

15. Pedersen M, Verspoor K, Jenkinson M, Law M, Abbott DF, Jackson GD. Artificial intelligence for clinical decision support in neurology. Brain Commun. 2020 Jul 1 [cited 2022 Jan 7];2(2). Available from: https://academic.oup.com/braincomms/article/doi/10.1093 /braincomms/fcaa096/5869431

16. Chen H, Sung JJY. Potentials of AI in medical image analysis in Gastroenterology and Hepatology. J Gastroenterol Hepatol [Internet]. 2021 Jan 15 [cited 2022 Jan 7];36(1):31–8. Available from: https://onlinelibrary.wiley.com/doi/10.1111/jgh.15327

17. Taghanaki SA, Abhishek K, Cohen JP, Cohen-Adad J, Hamarneh G. Deep Semantic Segmentation of Natural and Medical Images: A Review. 2019 Oct 16 [cited 2020 Nov 2]; Available from: http://arxiv.org/abs/1910.07655

18. Liu X, Song L, Liu S, Zhang Y. A review of deep-learning-based medical image segmentation methods. Sustain. 2021;13(3):1–29.

19. Nai YH, Teo BW, Tan NL, O'Doherty S, Stephenson MC, Thian YL, et al. Comparison of metrics for the evaluation of medical segmentations using prostate MRI dataset. Comput Biol Med. 2021 Jul 1;134:104477.

20. Kumar R V, Antony GM. A Review of Methods and Applications of the ROC Curve in Clinical Trials. Drug Inf J [Internet]. 2010 Nov 30 [cited 2022 Jan 8];44(6):659–71. Available from: http://link.springer.com/10.1177/009286151004400602

21. Hanley JA, McNeil BJ. The meaning and use of the area under a receiver operating characteristic (ROC) curve. Radiology [Internet]. 1982 [cited 2022 Jan 8];143(1):29–36. Available from: https://pubmed.ncbi.nlm.nih.gov/7063747/

22. Cohen J. A Coefficient of Agreement for Nominal Scales. Educ Psychol Meas [Internet]. 1960 Apr 2 [cited 2022 Jan 8];20(1):37–46. Available from: http://journals.sagepub.com/doi/10.1177/00131644600200 0104

23. Cohen's Kappa: what it is, when to use it, how to avoid pitfalls | KNIME [Internet]. [cited 2022 Jan 8]. Available from: https://www.knime.com/blog/cohens-kappa-an-overview

24. Delgado R, Tibau XA. Why Cohen's Kappa should be avoided as performance measure in classification. PLoS One [Internet]. 2019 Sep 1 [cited 2022 Jan 8];14(9):e0222916. Available from: https://doi.org/10.1371/journal.pone.0222916

25. Aydin OU, Taha AA, Hilbert A, Khalil AA, Galinovic I, Fiebach JB, et al. On the usage of average Hausdorff distance for segmentation performance assessment: hidden error when used for ranking. Eur Radiol Exp [Internet]. 2021 Dec 1 [cited 2022 Jan 8];5(1):4. Available from: https://eurradiolexp.springeropen.com/articles/10.1186/s4 1747-020-00200-2

26. Karimi D, Salcudean SE. Reducing the Hausdorff Distance in Medical Image Segmentation with Convolutional Neural Networks. IEEE Trans Med Imaging [Internet]. 2019 Apr 22 [cited 2022 Jan 8];39(2):499–513. Available from: http://arxiv.org/abs/1904.10030

27. GitHub [Internet]. [cited 2022 Jan 9]. Available from:

https://github.com/

28.  Zenodo - Research. Shared. [Internet]. [cited 2022 Jan 9]. Available from: https://zenodo.org/

29.  Müller D, Kramer F. MIScnn : a framework for medical image segmentation with convolutional neural networks and deep learning. BMC Med Imaging [Internet]. 2021 Jan 21;21:12. Available from: http://arxiv.org/abs/1910.09308

30.  Cheng J, Yang W, Huang M, Huang W, Jiang J, Zhou Y, et al. Retrieval of Brain Tumors by Adaptive Spatial Pooling and Fisher Vector Representation. Yap P-T, editor. PLoS One [Internet]. 2016 Jun 6 [cited 2022 Jan 9];11(6):e0157112. Available from: https://dx.plos.org/10.1371/journal.pone.0157112

31.  Cheng J, Huang W, Cao S, Yang R, Yang W, Yun Z, et al. Enhanced Performance of Brain Tumor Classification via Tumor Region Augmentation and Partition. Zhang D, editor. PLoS One [Internet]. 2015 Oct 8 [cited 2022 Jan 9];10(10):e0140381. Available from: https://dx.plos.org/10.1371/journal.pone.0140381

32.  Caicedo JC, Goodman A, Karhohs KW, Cimini BA, Ackerman J, Haghighi M, et al. Nucleus segmentation across imaging experiments: the 2018 Data Science Bowl. Nat Methods [Internet]. 2019 Dec 1 [cited 2022 Jan 9];16(12):1247–53. Available from: https://doi.org/10.1038/s41592-019-0612-7

33.  Ma J, Ge C, Wang Y, An X, Gao J, Yu Z, et al. COVID-19 CT Lung and Infection Segmentation Dataset [Internet]. Zenodo; 2020. Available from: https://doi.org/10.5281/zenodo.3757476

34.  Ma J, Wang Y, An X, Ge C, Yu Z, Chen J, et al. Towards Efficient COVID-19 CT Annotation: A Benchmark for Lung and Infection Segmentation [Internet]. 2020. p. 1–7. Available from: http://arxiv.org/abs/2004.12537